\documentclass[conference]{IEEEtran}
\IEEEoverridecommandlockouts
\usepackage{cite}
\usepackage{amsfonts}
\usepackage{booktabs} 
\usepackage{graphicx}
\usepackage{textcomp}
\usepackage{xcolor}
\usepackage{multirow} 
\usepackage{hyperref} 
\usepackage{amsmath}        
\usepackage{amssymb}        
\usepackage{bm}             

\usepackage{algorithm}      
\usepackage{algpseudocode}  

\usepackage{booktabs}       
\usepackage{array}          

\usepackage{graphicx}       
\usepackage{subcaption}     

\usepackage{xcolor}         
\usepackage{url}            

\def\BibTeX{{\rm B\kern-.05em{\sc i\kern-.025em b}\kern-.08em
    T\kern-.1667em\lower.7ex\hbox{E}\kern-.125emX}}
\begin{document}

\title{Dynamical Multimodal Fusion with Mixture-of-Experts for Localizations
\thanks{The work was supported by the National Natural Science Foundation of China under Grant 62331023 and 62394292, Zhejiang Provincial Science and Technology Plan Project under Grant 2024C01033, Zhejiang University Global Partnership Fund, and Fundamental Research Funds for the Central Universities under Grant No. 226-2024-00069.}
}

\author{
	\IEEEauthorblockN{Bohao Wang\IEEEauthorrefmark{1}\IEEEauthorrefmark{2}, Zitao Shuai\IEEEauthorrefmark{3}, Fenghao Zhu\IEEEauthorrefmark{1}, Chongwen Huang\IEEEauthorrefmark{1}\IEEEauthorrefmark{2}, Yongliang Shen\IEEEauthorrefmark{4}, }
    
    \IEEEauthorblockN{Zhaoyang Zhang\IEEEauthorrefmark{1}, Qianqian Yang\IEEEauthorrefmark{1}, Sami Muhaidat\IEEEauthorrefmark{5}, and M\'{e}rouane~Debbah\IEEEauthorrefmark{6}, \IEEEmembership{Fellow,~IEEE}}
	
	\IEEEauthorblockA{\IEEEauthorrefmark{1} College of Information Science and Electronic Engineering, Zhejiang University, 310027, Hangzhou, China}
	\IEEEauthorblockA{\IEEEauthorrefmark{2} State Key Laboratory of Integrated Service Networks, Xidian University, 710071, Xi'an, China}
    \IEEEauthorblockA{\IEEEauthorrefmark{3} Department of Electrical Engineering and Computer Science, University of Michigan, 48109, Ann Arbor, USA}
	
    \IEEEauthorblockA{\IEEEauthorrefmark{4} College of Computer Science and Technology, Zhejiang University, 310027, Hangzhou, China}
	\IEEEauthorblockA{\IEEEauthorrefmark{5} Computer and Communication Engineering, Khalifa University, P.O. Box: 127788, Abu Dhabi, UAE}
    \IEEEauthorblockA{\IEEEauthorrefmark{6} KU 6G Research Center, Khalifa University, P.O. Box: 127788, Abu Dhabi, UAE}
	
    }

\maketitle
\begin{abstract}
Multimodal fingerprinting is a crucial technique to sub-meter 6G integrated sensing and communications (ISAC) localization, but two hurdles block deployment: (i) the contribution each modality makes to the target position varies with the operating conditions such as carrier frequency, and (ii) spatial and fingerprint ambiguities markedly undermine localization accuracy, especially in non-line-of-sight (NLOS) scenarios.
To solve these problems, we introduce \textbf{SCADF-MoE}, a spatial-context aware dynamic fusion network built on a soft mixture-of-experts backbone.
SCADF-MoE first clusters neighboring  points into short trajectories to inject explicit spatial context.
Then, it adaptively fuses channel state information, angle of arrival profile, distance, and gain through its learnable MoE router, so that the most reliable cues dominate at each carrier band.
The fused representation is fed to a modality-task MoE that simultaneously regresses the coordinates of every vertex in the trajectory and its centroid, thereby exploiting inter-point correlations. 
Finally, an auxiliary maximum-mean-discrepancy loss enforces expert diversity and mitigates gradient interference, stabilizing multi-task training.
On three real urban layouts and three carrier bands (2.6, 6, 28 GHz), the model delivers consistent sub-meter MSE and halves unseen-NLOS error versus the best prior work.
To our knowledge, this is the first work that leverages large-scale multimodal MoE for frequency-robust ISAC localization.
\end{abstract}


\section{Introduction}
\begin{figure*}
    \centering
    \includegraphics[width=1\linewidth]{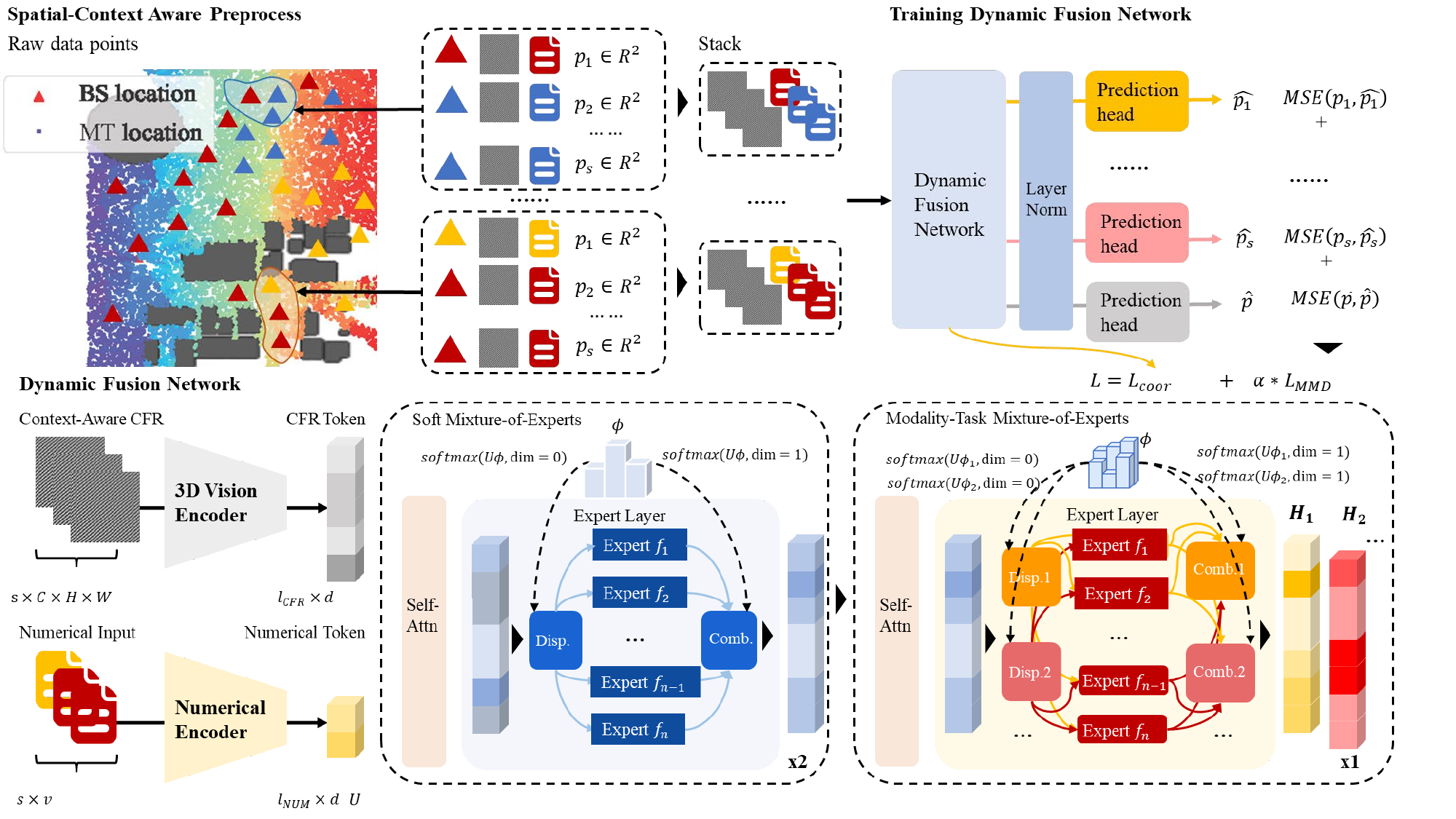}
    \caption{Overview of our SCADF-MoE method.}
    \label{fig:method2}
\end{figure*}

Sub–meter wireless localization is a keystone capability for 6G integrated sensing and communication (ISAC) networks, underpinning ultra-reliable beam management, digital-twin synchronization, autonomous driving, and smart-factory control~\cite{zhu2025wirelesslargeaimodel}.  
Achieving such accuracy in dense urban or industrial settings is notoriously hard because of rich multipath, frequent none-line-of-sight (NLOS) blockage, and carrier-dependent propagation effects severely degrade traditional model–based or single-modal fingerprinting solutions \cite{shahid2025largescaleaitelecomcharting,robustnetwork}.

Recent neural fingerprinting methods improve accuracy by fusing multiple modalities—e.g., channel-state-information (CSI) tensors with angle-of-arrival (AoA) profiles~\cite{10683036, sun2019fingerprint}.  
Yet two fundamental challenges persist:

(i) \textbf{Static fusion under changing conditions}. Most work model multi-modal interactions in a static way and therefore ignore that each modality’s usefulness can shift as operating conditions such as the carrier frequency, bandwidth, or propagation regime change. 
For instance, in spacious LOS areas distance and AoA remain stable and thus serve as trustworthy cues \cite{keusgen2022sub}. 
By contrast, in dense, metal-laden environments the channel frequency response (CFR) and the path-averaged large-scale gain remain dependable fingerprints because the CFR’s fine-grained amplitude-and-phase spectrum is largely preserved after heavy multipath scattering, and large-scale gain varies smoothly with geometry and therefore outlasts fast small-scale fading, together encoding richer location-specific detail\cite{chen2016achieving}.  
Overlooking such frequency variability prevents a system from exploiting the strongest cue per band and cripples its ability to generalize when the carrier changes \cite{wudynamic}.  
This motivates our first research question:  \textbf{How can we design a fusion strategy that dynamically adapts to different frequency bands, delivering consistently accurate multimodal localization?}

(ii) \textbf{Spatial-fingerprint ambiguity}. Wireless fingerprints recorded at different positions often look deceptively alike, especially under NLOS, because multipath and blocking objects damage the mapping between signal space and physical space \cite{leitinger2019belief,wang2024multi}.  
Consequently distant users may be projected onto neighboring grid points, while adjacent references lack distinctive features \cite{sun2013moloc}.  
Existing fixes, such as hand-crafted features \cite{del2019localization}, data augmentation \cite{li2021wireless1}, domain-adversarial learning \cite{li2021wireless}, or auxiliary sensors \cite{petrushin2006multiple}, either need extra hardware or still treat samples independently, missing the rich \emph{spatial correlations} among nearby points.  
This leads to our second research question: \textbf{How can we exploit spatial context within multimodal wireless data itself to resolve spatial and fingerprint ambiguity and sustain high accuracy?}

In practice, neighbouring locations exhibit highly correlated channel signatures while still being distinguishable at a global scale \cite{sun2013moloc}, suggesting that localization should be formulated as a \emph{multi-task} problem that jointly regresses the coordinates of all points in a short trajectory.  
SCADF-MoE (i) uses \emph{soft MoE routers} to re-weight CSI, AoA, and related cues on a \emph{per-sample} basis, automatically adapting to carrier-frequency changes, and (ii) jointly outputs the coordinates of each point in a short neighborhood trajectory together with its centroid, so that a single ambiguous fingerprints becomes a discriminative multi-task signal.
We ground the model with an extended WAIR-D dataset covering three carrier bands, and convert isolated samples into locally correlated yet globally distinctive trajectories via radius-based clustering.
Next, a band-adaptive soft MoE fusion then assigns input-conditioned modality weights, while a subsequent modality-task MoE simultaneously regresses all trajectory points, explicitly exploiting shared spatial structure.
To ensure complementary expert specialization and stable gradients, we attach a lightweight maximum-mean-discrepancy regularize.
Evaluated on three urban layouts and all carrier bands, SCADF-MoE cuts MSE by \textbf{63\%} relative to the strongest baseline and halves unseen-NLOS error, consistently yielding sub-meter accuracy.

\section{Problem Formulation and Spatial-Context Data}
\label{sec:data}

\subsection{Multimodal Wireless Fingerprint}

We employ a high-fidelity ray-tracer to synthesize the path-wise channel impulse response (CIR) at every location.
Taking the Fourier transform of CIR on each $f_l$ yields
\begin{equation}
\label{eq:CFR}
H(f_l)=\sum_{i=1}^{N}a_i\,e^{j\phi_i}\,
       e^{-j2\pi f_l\tau_i},
\end{equation}
where $f_l=f_c+\frac{l}{T_{\mathrm{s}}N_c}$ $(l=0,\dots,N_c\!-\!1)$. 
Stacking \eqref{eq:CFR} over $l$ forms the CFR vector
$\mathbf{h}\!=\![H(f_0),\dots,H(f_{N_c-1})]^{\!T}\in\mathbb{C}^{N_c}$.
From the CFR we form a three-channel tensor
\(\mathbf I(\mathbf p)\in\mathbb R^{3\times N_c}\)
(\emph{real}, \emph{imag}, \emph{magnitude}) and extract four geometry-aware scalars—
dominant AoD, AoA, propagation distance, and large-scale gain— as \(\mathbf v(\mathbf p) =
\bigl[
\theta^{(\mathrm{t})},\;
\theta^{(\mathrm{r})},\;
d,\;
g
\bigr]^{\!\top}\in\mathbb R^{4}\).
The complete multimodal fingerprint at position
\(\mathbf p\in\mathbb R^{2}\) is therefore
\begin{equation}
\label{eq:fingerprint}
\mathcal F(\mathbf p)=\bigl\{
      \underbrace{\mathbf I(\mathbf p)}_{\text{CFR image}},\;
      \underbrace{\mathbf v(\mathbf p)}_{\text{geometry \& gain}}
    \bigr\}.
\end{equation}

\subsection{Trajectory Sampling from Spatial Clusters}
\label{sec:traj_sampling}
To inject spatial context while avoiding redundancies, we
convert isolated fingerprints \(\mathcal F(\mathbf p_i)\) into
short neighbourhood trajectories through three steps:
radius-based clustering, train–test partition, and trajectory
generation.

Firstly, for each anchor \(\mathbf p_i\) we form its $r$-ball
\(
\mathcal N_r(\mathbf p_i)=
 \{\mathbf p_j\mid\lVert\mathbf p_i-\mathbf p_j\rVert_2\le r\}
\)
and merge overlapping balls.
The radius \(r^{\star}\) is chosen as the smallest value for which the
median cluster size reaches a preset threshold \(n_{\min}\) so
that at least half of the clusters contain sufficient neighbours:
\[
r^{\star}
  =\min\bigl\{r>0\mid\operatorname{median}_u\lvert\mathcal C_u(r)\rvert
               \ge n_{\min}\bigr\}.
\]
Using \(r=r^{\star}\) yields \(U\) disjoint clusters
\(\mathcal C_1,\dots,\mathcal C_U\).

Then, each cluster is divided into $70\%$ training and $30\%$
testing points.  
From the training points subset, we draw length-\(s\) trajectories \(\tau=(\mathbf p_{1},\dots,\mathbf p_{s})\)
by uniform sampling.
To fairly evaluate generalization, a test trajectory is accepted only if it contains at least one point that never appears in training. 
This simple rule guarantees every test sample includes an unseen location while preserving comparable path lengths across splits.


\subsection{Multimodal Feature Assembly}
Each trajectory is converted into two aligned input branches:

\begin{enumerate}
\item \textbf{Numeric branch}—concatenate the four scalar
      descriptors of every vertex:
      \begin{equation}
        \label{eq:vec}
        \mathbf V^{\text{Num}}_{\tau}
        =[\theta^{(\mathrm t)}_{1},\theta^{(\mathrm r)}_{1},d_1,g_1,
          \dots,
          \theta^{(\mathrm t)}_{s},\theta^{(\mathrm r)}_{s},d_s,g_s]^\top
        \in\mathbb R^{4s}.
      \end{equation}

\item \textbf{CFR branch}—stack the three-channel CFR tensors
      of all vertices into a mini‐sequence:
      \begin{equation}
        \label{eq:cfr}
        \mathbf V^{\text{CFR}}_{\tau}
          =[\mathbf I(\mathbf p_1);\dots;\mathbf I(\mathbf p_s)]
          \in\mathbb R^{s\times3\times H\times W}.
      \end{equation}
\end{enumerate}

Together, \((\mathbf V^{\text{Num}}_{\tau},\mathbf V^{\text{CFR}}_{\tau})\)
and the target coordinate set
\(
\mathbf P_{\tau}=[\mathbf p_{1},\dots,\mathbf p_{s},\bar{\mathbf p}]
\) form one training example, 
where $\bar{\mathbf{p}}
   =\frac{1}{s}\sum_{i=1}^{s}\mathbf{p}_{i}\in\mathbb{R}^{2}$
. Converting isolated points into trajectory-level samples thus injects spatial context while keeping the input dimensionality modest, which is an essential ingredient for the downstream SCADF-MoE network.





\section{Proposed SCADF-MoE Architecture}
\label{sec:scadf}

SCADF-MoE is designed to adaptively fuse heterogeneous
modalities and jointly localize a short trajectory of neighboring points.
Fig.~\ref{fig:method2} gives a block-level view.

\subsection{End-to-End Pipeline}
\label{sec:pipeline}
\begin{algorithm}[t]
  \caption{SCADF Training Pipeline}
  \label{alg:scadf_short}
  \begin{algorithmic}[1]
    \Require dataset $\mathcal D$, epochs $E$, learning rate $\eta$, MMD weight $\alpha$
    \State Initialize encoders $\{\mathcal E_i\}$, Soft MoE stack $\mathcal S$, Modality-Task MoE $\mathcal M$, optimizer Adam($\theta$)
    \For{$e=1$ \textbf{to} $E$}
      \For{minibatch $(\mathbf V^{\text{Num}}_\tau,\mathbf V^{\text{CFR}}_\tau,\mathbf P_\tau)\!\in\!\mathcal D$}
        \State \textbf{Encoding:}\; $X_i\!\leftarrow\!\mathcal E_i(\mathbf V^{(i)}_\tau)$  \hfill($i\!=\!\{\text{Num},\text{CFR}\}$)
        \State \textbf{Fusion:}\; $Z\!\leftarrow\!\mathcal S(X_{\text{Num}},X_{\text{CFR}})$
        \State \textbf{Multi-task heads:}\; $\{\hat{\mathbf p}_k\}_{k=1}^{s},\hat{\bar{\mathbf p}}\!\leftarrow\!\mathcal M(Z)$
        \State \textbf{Loss:}\; $\mathcal L=\mathcal L_{\text{coord}}+\alpha\mathcal L_{\text{MMD}}$
        \State \textbf{Update:}\; $\theta\leftarrow\theta-\eta\nabla_\theta\mathcal L$
      \EndFor
    \EndFor
  \end{algorithmic}
\end{algorithm}

\textbf{Step\,1: Tokenization.}
A CFR encoder converts the $\mathbf I(\mathbf p)$ into a sequence of
$l_\text{CFR}$ tokens by a depth-wise convolution followed by a shallow 1-D CNN.
Positional embeddings are then added to capture frequency order.
A separate numeric encoder projects the feature vector
$\mathbf v(\mathbf p)$ into an equal-length token steam via a lightweight MLP.
Throughout our experiments, the two token streams share the same length and hidden dimension.

\textbf{Step\,2: Soft MoE Fusion.}
The two token streams are concatenated and forwarded through a sequence of soft MoE blocks, each of which contains $n$ experts, to achieve dynamic integration of effective information from respective modalities under the guidance of the routing matrix. 
Unlike sparse MoE variants that drop most tokens, we choose a soft router for two reasons:
(i)~wireless fingerprints are low-dimensional, so a dense mixture incurs little extra floating point operations per second (FLOP); 
(ii)~soft routing yields smooth gradients, leading to faster convergence.

\textbf{Step\,3: Modality-Task MoE.}
The $M$ fused tokens output by the
last soft MoE layer are passed to a second MoE whose
\((s+1)\) routers are \emph{task specific}.
Router $k$ (for point or the centroid) generates dispatch weights \(
D^k
   =\mathrm{softmax}_{\text{row}}\bigl(\mathbf U\Phi_{\mathrm{task},k}\bigr)
\) and combination weights \(
C^k
   =\mathrm{softmax}_{\text{col}}\bigl(\mathbf U\Phi_{\mathrm{task},k}\bigr).
\)
exactly.
In practice we re-use the same $n$ experts but employ independent 1-D task embeddings
$\mathrm{TE}^k\!\in\!\mathbb R^{d}$ to bias the expert’s MLP toward
task-relevant features.
A single linear head maps the resulting embeddings to coordinates. 
The algorithm for the end-to-end pipeline is shown as Algorithm \ref{alg:scadf_short}.


\subsection{Dynamic Multimodal Fusion via Soft MoE}
\label{sec:softmoe}
In this section, we will introduce the soft MoE network architecture. Each soft MoE block comprises an attention block, an expert layer with $n$ expert functions $\{f_i\}^n_{i=1}$, and a learnable routing matrix $\Phi$. 

For each input sample, we first encode its \(m\) modalities into token embeddings and concatenate them,
\[
\mathbf X\in\mathbb R^{M\times d},\quad
M=\sum_{i=1}^{m} l_i,
\]
where \(l_i\) is the token count of modality~\(i\) and \(d\) the embedding dimension.
Throughout this work we use two modalities: the CFR tensor \(V_{\tau}^{\mathrm{CFR}}\) and the numerical vector \(V_{\tau}^{\mathrm{Num}}\).
The concatenated tokens are processed by a multi-head self-attention (MSA) block,
\[
\mathbf U = \mathbf X + \mathrm{MSA}(\mathbf X),
\]
after which the standard Transformer feed-forward (MLP) layer is replaced by a MoE layer, providing sample-adaptive, modality-aware fusion. The expert layer dynamically fuses information from multi-modalities by creating sample-dependent weight pathways for each input. This helps us balance the information from CFR and numeric measurements on a sample-to-sample basis, learning changing quality and relevance of different modalities across heterogeneous environments. The routing process is scheduled by a learnable matrix $\Phi$, which will be used to generate input-conditioned expert-token routing matrices. Specifically, the product of $\Phi$ and input is normalized appropriately to compute both the input-to-expert dispatch weights and expert-to-output combine weights.


To route the input tokens \(\mathbf{U}\in\mathbb{R}^{M\times d}\) to \(n\) experts, we multiply a learnable routing matrix $\phi\in\mathbb{R}^{M\times n}$ with the input tokens to obtain a dispatch weight matrix \(\mathbf{D}\in\mathbb{R}^{M\times n}\) via:
\[
D_{ij}
= \frac{\exp\bigl((\mathbf{U}\,\Phi)_{ij}\bigr)}
       {\sum_{i'=1}^{M}\exp\bigl((\mathbf{U}\,\Phi)_{i'j}\bigr)},
\]
where \(\Phi\in\mathbb{R}^{d\times n}\) is the learnable routing tensor. The expert inputs are then linearly aggregated as:
\[
\widetilde{\mathbf{U}}
= \mathbf{D}^\top \mathbf{U},
\quad
\widetilde{\mathbf{U}}\in\mathbb{R}^{n\times d}.
\]

Each aggregated vector \(\widetilde{\mathbf{U}}_i\in\mathbb{R}^d\) is then processed by expert \(f_k\), the expert’s output is
\[
\widetilde{\mathbf{Y}}_i = f_k\bigl(\widetilde{\mathbf{U}}_i\bigr).
\]

In our implementation, each \(f_k\) is realized as a lightweight MLP block.

To gather the output of each expert, we multiply the routing matrix $\Phi$ with $U$ to obtain the combined weights \(\mathbf{C}\in\mathbb{R}^{M\times n}\):
\[
C_{ij}
=\frac{\exp\bigl((\mathbf{U}\,\Phi)_{ij}\bigr)}
     {\sum_{j'=1}^{n}\exp\bigl((\mathbf{U}\,\Phi)_{ij'}\bigr)},
\quad
\mathbf{Y} = \mathbf{C}\,\widetilde{\mathbf{Y}},
\]
where \(\mathbf{Y}\in\mathbb{R}^{M\times d}\) is the expert layer's output. Then the output is added residually by:
\[
\mathbf{Z} = \mathbf{U} + \mathbf{Y}.
\]

\subsection{Multi-Task Localization via Modality-Task MoE}
\label{sec:modtask}
After multimodal fusion, the token matrix
\(\mathbf Z\!\in\!\mathbb R^{M\times d}\) is fed to a
Modality-Task MoE that produces
\(K=s+1\) outputs—the \(s\) vertex coordinates of a trajectory and its centroid.
The block inherits the attention–expert structure of Soft-MoE, but replaces the \emph{single} router with a
\emph{task-specific} router family
\(\Phi_{\text{task}}\in\mathbb R^{d\times n\times K}\).

A multi-head self-attention layer first yields
\(\mathbf U=\mathbf Z+\text{MSA}(\mathbf Z)\).
For each task \(k\in\{1,\dots,K\}\) we select slice
\(\Phi_{\text{task},k}\!\in\!\mathbb R^{d\times n}\) and compute
\[
D^k_{ij}
= \frac{\exp\bigl((\mathbf U\,\Phi_{\mathrm{task},..k})_{ij}\bigr)}
       {\sum_{i'=1}^M\exp\bigl((\mathbf U\,\Phi_{\mathrm{task},..k})_{i'j}\bigr)}.
\]

Then for each task $k$, its corresponding dispatch matrix assigns tokens \(\widetilde U^k_{j}\) to each expert \(j\) as \(
\widetilde{\mathbf U}^{k}=D^{k\top}\mathbf U,
\). Notably, the \(n\) expert networks themselves remain shared across all tasks, with each expert processing the token sets dispatched by every task router differently.

To capture fine-grained, task-specific cues, we add the learnable task embedding \(\mathrm{TE}^k\) to the weighted averaged tokens for each point before expert processing:
\[
\widetilde{\mathbf Y}^{k}_{j}
   =f_j\!\bigl(\widetilde{\mathbf U}^{k}_{j}+\mathrm{TE}^{k}\bigr).
\]

Then, to aggregate the outputs of each expert, for each task \(k\), we obtain a task-specific combine matrix \(C^k\in\mathbb{R}^{M\times n}\) following:

\[
C^k_{ij}
= \frac{\exp\bigl((\mathbf U\,\Phi_{\mathrm{task,..k}})_{ij}\bigr)}
       {\sum_{j'=1}^n\exp\bigl((\mathbf U\,\Phi_{\mathrm{task,..k}})_{ij'}\bigr)},
\quad
Y^k_i = \sum_{j} C^k_{ij}\,\widetilde Y^k_{j},
\]
yielding \(\mathbf Y^1, ..., \mathbf Y^{s+1} \in\mathbb R^{M\times d}\).  A residual connection \(\mathbf H=\mathbf U+\mathbf Y\) is also applied.

A lightweight linear head then maps each normalized embedding to a 2D coordinate:
\[
\hat p_k = W_k\,\widetilde H^k + b_k,\quad
\hat{\bar p} = W_{\mathrm{avg}}\,\widetilde{\bar H} + b_{\mathrm{avg}}.
\]
We optimize the mean-squared error (MSE) over all \(s+1\) coordinates' predictions and their centroid:
\[
\mathcal L_{\mathrm{coord}}
= \frac{1}{s+1}\Bigl(\sum_{k=1}^{s}\|\hat p_k - p_k\|_2^2
\;+\;\|\hat{\bar p} - \bar p\|_2^2\Bigr),
\quad
\bar p = \frac{1}{s}\sum_{k=1}^{s}p_k.
\]

This design yields three advantages: 
(i) The centroid head captures the \emph{global} geometry of the
trajectory and regularizes the $s$ point heads;
(ii) experts are shared across tasks, so features useful to several
points (e.g., NLOS cues) are learned once, improving data efficiency;
(iii) task-specific routers allow each point to activate a different
expert mixture, yielding sample- and task-adaptive localization.




\vspace{1mm}
\subsection{MMD-Regularized Expert Diversity}
Although assigning a private router to every task encourages a degree
of specialization, there is no guarantee that different routers will
select different experts.  
In practice, several tasks can collapse onto the same expert
mixture because the primary loss cares only about prediction accuracy,
not about the diversity of the computational paths.  
Such collapse wastes model capacity and re-introduces gradient
interference, as the shared experts must now accommodate possibly
conflicting objectives.  
To avert this problem we add an MMD term that enlarges the pairwise distance
between the dispatch and combine matrices generated by different
routers, thereby forcing tasks to explore complementary expert
subsets while still permitting overlap when it is genuinely useful.

\[
\mathcal L_{\text{MMD}}
= \frac{2}{K(K-1)}
  \textstyle\sum_{p<q}
  \bigl[\mathrm{MMD}^2(D^{p},D^{q})
       +\mathrm{MMD}^2(C^{p},C^{q})\bigr],
\]
where
\(
\mathrm{MMD}^2(A,B)
  = \frac{1}{M^2}\!\sum_{i,j}
     \!k(A_{i,:},A_{j,:})
  + \frac{1}{M^2}\!\sum_{i,j}
     \!k(B_{i,:},B_{j,:})
  - \frac{2}{M^2}\!\sum_{i,j}
     \!k(A_{i,:},B_{j,:}),
\)
\(k(\mathbf x,\mathbf y)=\exp\bigl(-\lVert\mathbf x-\mathbf y\rVert^2/2\sigma^2\bigr)\), 
the kernel is
\(
\mathcal{K}(x,y)
= \exp\!\Bigl(-\frac{\lVert x - y\rVert^2}{2\sigma^2}\Bigr).
\)
Thus the overall training objective of our method could be written as: 
\begin{equation}
\label{final loss}
\mathcal L = \mathcal L_{coord} + \alpha \mathcal L_{\mathrm{MMD}},    
\end{equation}
where we adopt \(\sigma=\sqrt{d/2}\) and set the weight
\(\alpha\), which empirically balances diversity and accuracy.

\section{Numerical Results}
\label{sec:exp}

We assess SCADF-MoE from four angles:
\emph{(i)} overall accuracy,
\emph{(ii)} generalization to unseen \emph{frequency} bands,
\emph{(iii)} sensitivity to key hyper-parameters,
and \emph{(iv)} ablation of model components.
All results average five independent runs; 95\% confidence
intervals are within the marker size.

\subsection{Experimental Setup}
\textbf{Hardware \& software.}
Training is performed on an Ubuntu\,24.04 server with an AMD EPYC-9654
CPU and four NVIDIA A100 (40 GB) GPUs.
Code is implemented in PyTorch 2.3 with CUDA 12.4 and cuDNN 9.1.
We use AdamW ($\text{lr}=1\!\times\!10^{-4}$, weight-decay
$5\!\times\!10^{-5}$, batch 64) and cosine learning-rate decay.

\textbf{Ray-tracing data.}
We extend the WAIR-D mmWave-MIMO corpus~\cite{huangfu2022wair}
to three carrier bands—2.6, 6, and 28 GHz—across the
urban scenes \texttt{00743}, \texttt{01105}, and \texttt{08927}.
Each site provides LOS/NLOS labels and full multipath descriptors.
Neighbourhood clustering (radius 2 m, trajectory length $s=5$)
inflates the data to \(\sim1.5\) M trajectory samples,
adequate for MoE training.
Other parameters are fixed (50 MHz bandwidth, 64 subcarriers,
BS array $1{\times}64$, UE isotropic).
Key simulation parameters are listed in Table~\ref{tab:dataset}.

\begin{table}[t]
  \centering
  \caption{Key simulation parameters.}
  \label{tab:dataset}
  \renewcommand{\arraystretch}{1.05}
  \begin{tabular}{@{}ll@{}}
    \toprule
    \textbf{Parameter} & \textbf{Value} \\ \midrule
    Carrier bands $f_c$ & \{2.6, 6, 28\}\,GHz \\
    Bandwidth           & 50 MHz, 64 subcarriers \\
    BS / UE arrays      & $1{\times}64$ ULA / single-element \\
    Trajectory length $s$ & 5 points, radius 2m \\
    Scenes (\# pts)     & 00743, 01105, 08927  \\ 
    Regularization weight $\alpha$    & 0.1 \\ \bottomrule
  \end{tabular}
\vspace{-2mm}
\end{table}

\textbf{Baselines.}
We compare against four fusion strategies—\textit{Concat}, \textit{FullCon} \cite{arnold2018deep}, \textit{TransFusion} \cite{xu2024swin}, and \textit{SoftMoE} \cite{puigcerver2023sparse}—all augmented with the same spatial-context sampling, unimodal encoders, and coordinate-regression head to ensure a fair comparison.
Their only difference lies in \emph{how} the CFR and numeric branches are merged:

\begin{itemize}
\item \textbf{Concat:} mean-pool each modality and simply concatenate the two fixed-length vectors—no learnable fusion weights, hence fully \emph{static}.
\item \textbf{FullCon:} replaces the hard concat with a single fully connected layer, giving a globally learned but sample-invariant fusion matrix.
\item \textbf{TransFusion:} inserts a Transformer encoder; self-attention provides dense, token-level interactions that are data-dependent yet use one shared set of parameters for all samples.
\item \textbf{SoftMoE:} adopts a soft Mixture-of-Experts router; tokens are routed through multiple experts with sample-specific soft gates, but experts are shared across tasks, and no spatial-context awareness is built in.
\end{itemize}

Thus, while all four baselines employ identical encoders and heads, they progress from \emph{static} (Concat) to \emph{globally learned} (FullCon) to \emph{dense adaptive} (TransFusion) to \emph{sparsely adaptive} (SoftMoE) fusion, highlighting the added value of SCADF-MoE’s frequency-aware \emph{and} task-aware expert routing.

\textbf{Metric.}
Localization error is measured by the MSE
\(\frac1N\sum_{i}\lVert\hat{\mathbf p}_i-\mathbf p_i\rVert_2^{2}\).

\subsection{Effect of Spatial-Context Sampling}
\begin{figure}
    \centering
    \includegraphics[width=\linewidth]{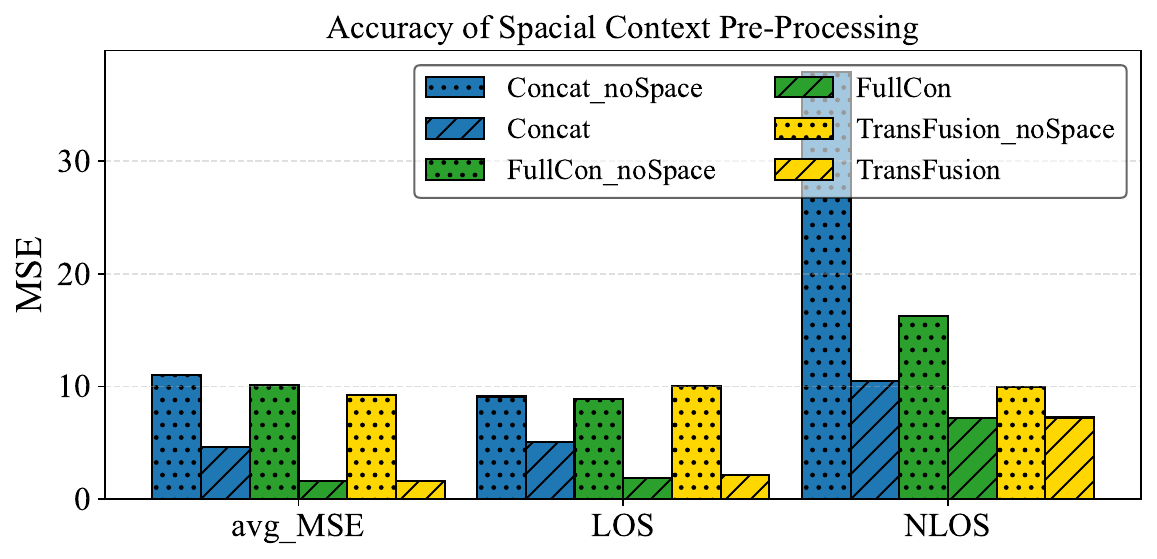}
    \caption{Effectiveness of spatial context pre-processing.}
    \label{fig:spatialContext}
\end{figure}
We first evaluate the effectiveness of our pre-processing techniques. 
Fig.~\ref{fig:spatialContext} plots LOS/NLOS histograms with and without the proposed neighbourhood sampling.
Removing spatial context (\texttt{\_noSpace}) rises the MSE by roughly \(3\!\times\)–\(5\!\times\) and hurts NLOS the most, confirming that short trajectories supply
essential geometric cues.

\subsection{Main Result: Frequency Heterogeneity}
\begin{table}[t]
  \centering
  \caption{Scene-wise overall and NLOS$_u$ errors for antenna-array mix (\(\times10^{-2}\,\mathrm{m}^{2}\)).}
  \label{tab:scene_ovr_nlosu}
  \renewcommand{\arraystretch}{1.15}
  \setlength{\tabcolsep}{6pt}
  \begin{tabular}{lcccccc}
    \toprule
      \multirow{2}{*}{\textbf{Method}} &
      \multicolumn{2}{c}{\shortstack{00743\\(Dense-Urban Campus)}} &
      \multicolumn{2}{c}{\shortstack{01105\\(Suburban Strip)}} &
      \multicolumn{2}{c}{\shortstack{08927\\(Urban Canyon)}} \\
      \cmidrule(lr){2-3}\cmidrule(lr){4-5}\cmidrule(lr){6-7}
      & \textbf{Ovr.} & \textbf{NLOS}$_u$
      & \textbf{Ovr.} & \textbf{NLOS}$_u$
      & \textbf{Ovr.} & \textbf{NLOS}$_u$ \\
    \midrule
      Concat        & 5.82 & 10.6 & 5.94 & 11.7 & 5.68 & 11.5 \\
      FullCon       & 2.02 &  4.45 & 1.80 &  4.00 & 1.93 &  4.20 \\
      TransFusion   & 2.53 &  4.90 & 1.93 &  4.30 & 2.05 &  4.45 \\
      SoftMoE       & 1.40 &  3.02 & 1.37 &  3.00 & 1.45 &  3.08 \\
      \textbf{SCADF-MoE}
                    & \textbf{0.34} & \textbf{1.20}
                    & \textbf{0.53} & \textbf{1.12}
                    & \textbf{0.56} & \textbf{1.16} \\
    \bottomrule
  \end{tabular}
\end{table}


Two evaluation regimes are considered:

\begin{itemize}
\item \textbf{Mix}: training and test batches draw uniformly from
      the three bands (in-domain).
\item \textbf{OOD}: the model trains on \{2.6, 28\}\,GHz and is
      \emph{only} tested at the previously unseen 6 GHz band.
\end{itemize}

\begin{figure}[t]
 \centering
 \includegraphics[width=\linewidth]{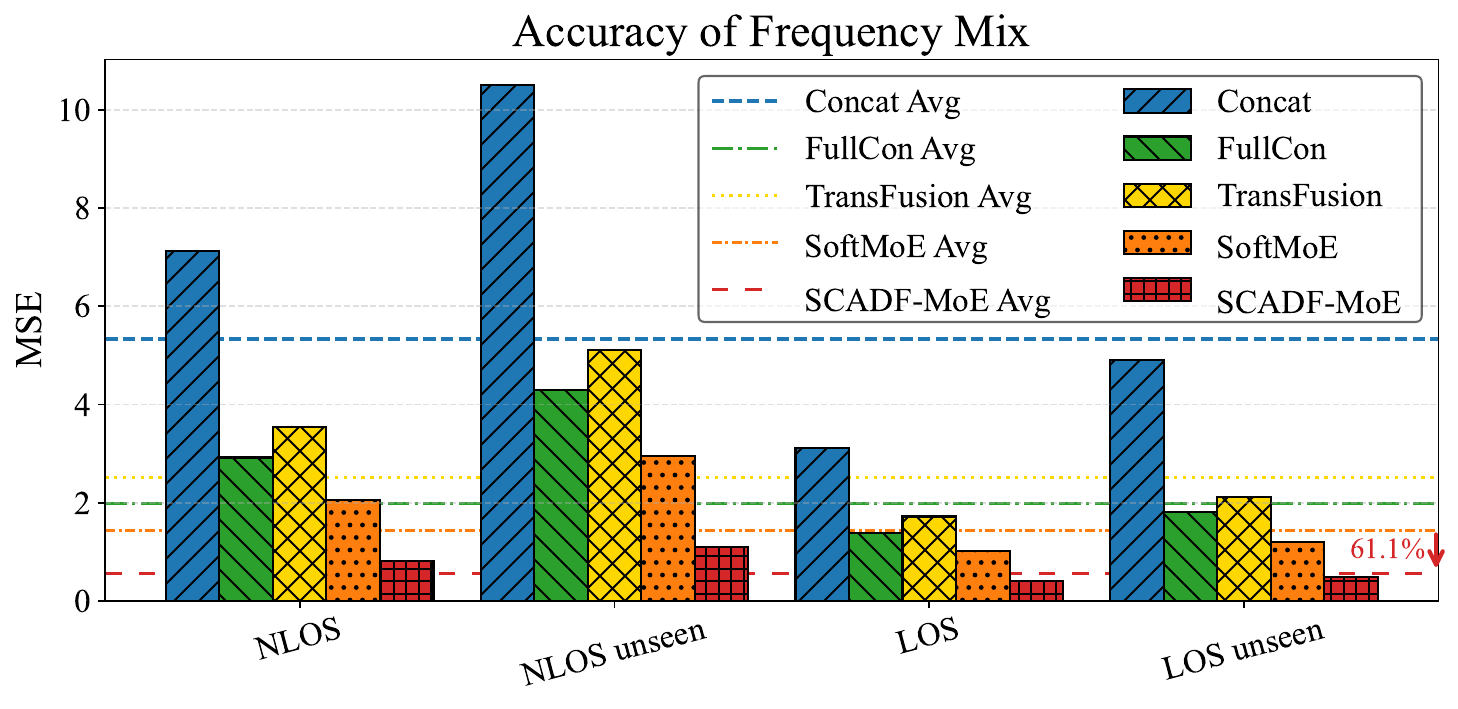}\\[-2pt]
 \includegraphics[width=\linewidth]{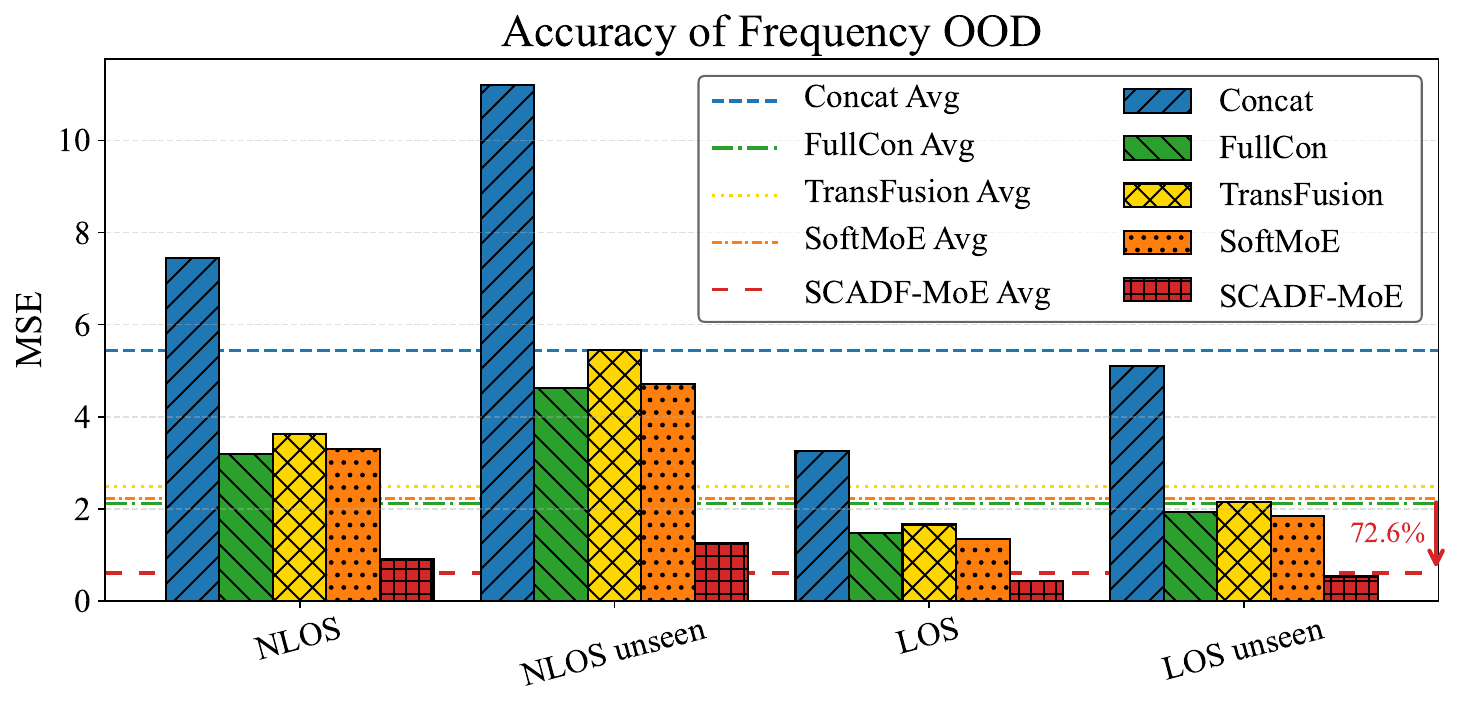}
 \caption{MSE across carrier bands: (top) Mix, (bottom) OOD.}
 \label{fig:freq}
\end{figure}

In the \textbf{Mix} setting (top of Fig.\,\ref{fig:freq}), evaluated on the \texttt{00743} scene, 
SCADF-MoE attains an overall \(\mathrm{MSE}=0.56~\text{m}^2\), \textbf{61\%} lower than the strongest baseline (SoftMoE) and well below the 1m\(^2\) target.
When switching to the harder \textbf{OOD} split (bottom of Fig.\,\ref{fig:freq}), the error remains below 1m\(^2\) for SCADF-MoE, whereas all competitors deteriorate sharply, confirming the value of frequency-adaptive expert routing.

Extending the evaluation of \textbf{Mix} setting to the other two urban layouts (\texttt{01105} and \texttt{08927}) yields the same trend: across \emph{all three} scenes and carrier bands, SCADF-MoE consistently delivers sub-meter accuracy and outperforms the best baseline as shown in the Table \ref{tab:scene_ovr_nlosu}, underscoring the combined importance of frequency-adaptive multimodal fusion and spatial-context modelling.





\subsection{Ablation Study}

\begin{figure}
    \centering
    \includegraphics[width=\linewidth]{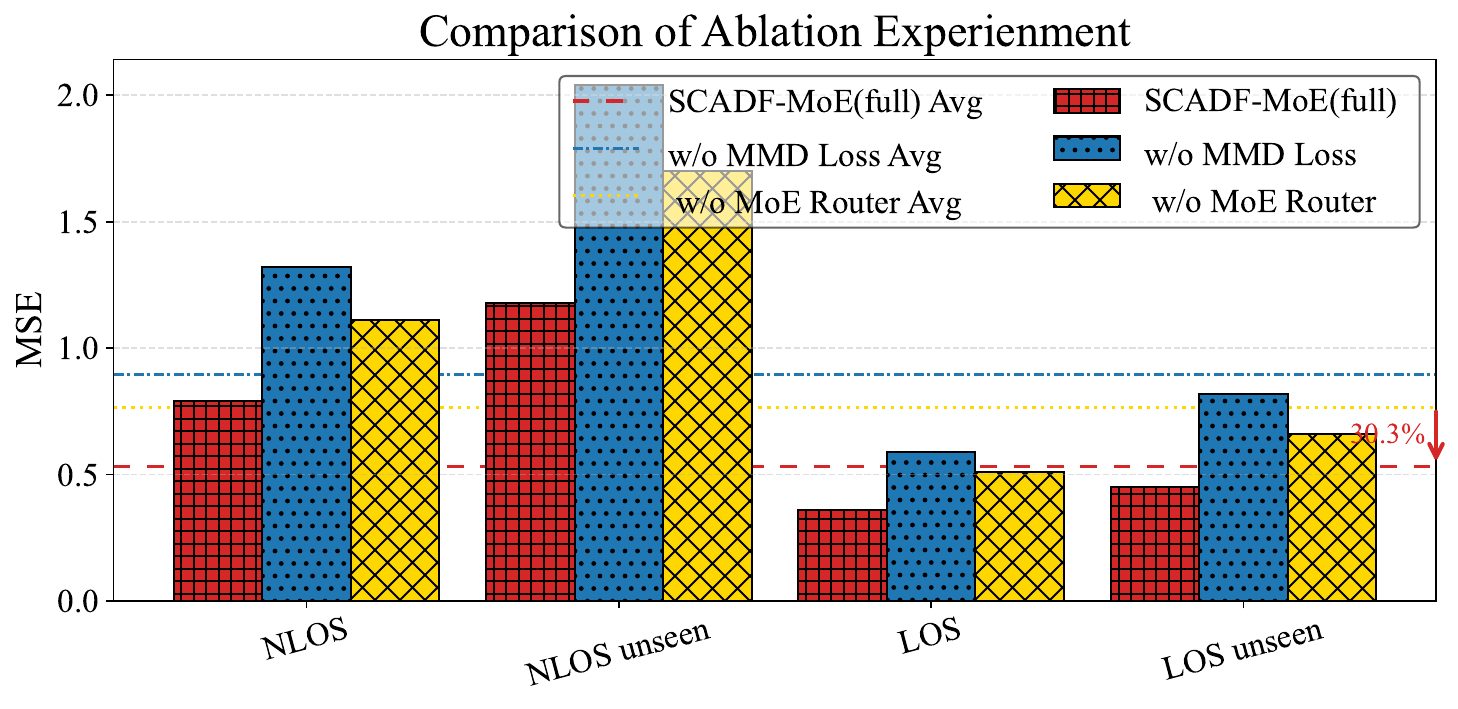}
    \caption{Ablation experiment of SCADF-MoE in antenna mix.}
    \label{fig:ablation}
    \vspace{-1mm}
\end{figure}

Fig.~\ref{fig:ablation} shows that removing either spatial context
or dynamic routing more than doubles the error—especially in NLOS—while collapsing to a single expert yields similar degradation.
Hence both adaptive fusion and trajectory supervision are indispensable.

\medskip
\noindent

\section{Conclusion}\label{sec:conclusion}
We presented SCADF-MoE, the first spatial-context-aware MoE framework for 6G high-precision localization.  
By converting isolated ray-tracing samples into short neighbourhood trajectories and letting soft-gated experts fuse CFR and geometric cues on a per-sample, per-task basis, the proposed network resolves both frequency-dependent modality drift and NLOS fingerprint and spatial  ambiguity.  
Extensive simulations on three urban layouts and three carrier bands verify its effectiveness: SCADF-MoE cuts average MSE by 63\% and unseen-NLOS error by 55\% relative to the strongest prior art, consistently delivering sub-meter accuracy.  
These results underscore the promise of MoE-driven multimodal fusion as a scalable solution for 6G ISAC positioning.  
Future work will integrate real-time measurement, introduce more heterogeneous operating conditions, and investigate model-compression techniques so that the approach can be deployed on edge hardware.  
In summary, this study provides a precise and robust new blueprint for carrier-diverse, sub-meter localization in next-generation wireless networks.

\small
\bibliographystyle{IEEEtran}
\bibliography{bib}
\vspace{12pt}

\end{document}